\documentclass[11pt]{article}
\usepackage{mymoriond,graphicx}

\def\be{\begin{equation}}
\def\ee{\end{equation}}
\def\ba{\begin{eqnarray}}
\def\ea{\end{eqnarray}}
\begin{document}
\parbox[][4cm][t]{\textwidth}{\begin{flushright}
LU TP 03-19\\hep-ph/0304284\\April 2003\end{flushright}}
\title{CHIRAL PERTURBATION THEORY AT TWO LOOPS AND THE MEASUREMENT OF
$V_{us}$\footnote{Talk given at 38th Rencontres de Moriond on QCD and
High-Energy Hadronic Interactions, Les Arcs, Savoie, France, 22-29 Mar 2003.} }

\author{ J.~BIJNENS}

\address{Department of Theoretical Physics 2, Lund University,\\
S\"olvegatan 14A, S 22362 Lund, Sweden}

\maketitle\abstracts{I give an overview of the calculations done in
three-flavour Chiral perturbation theory
at next-to-next-to-leading order with an emphasis on those relevant
for an improvement in the accuracy of the measurement of $V_{us}$.
It is pointed out that all needed low energy constants can be obtained
from experiment via the scalar form-factor in $K_{\ell3}$ decays.
}

\section{Introduction}
\label{intro}

The precise determination of the elements of the Cabibbo-Kobayashi-Maskawa
matrix (CKM) is an important part of the study of the flavour sector of the
standard model. A recent overview can be found in the proceedings of
the CERN CKM workshop.~\cite{CKM}
In this talk I will concentrate on the theory behind the measurement
of $V_{us}$ from $K_{\ell3}$ ($K\to\pi\ell\nu$) decays and in particular
on the recent work of P.~Talavera and myself on the two-loop calculation
and the determination of the relevant low-energy constants.~\cite{BT2}

I first give a short overview of Chiral Perturbation Theory (ChPT) and
the relevant two-loop calculations which have been performed to date. Then I
discuss $K_{\ell3}$ and the present situation for the theory. I also
include a short discussion of the validity of the linear approximation of
the form factors normally used in the data analysis. I then proceed to the
results from ChPT~\cite{BT2} which can be summarized as follows. The curvatures
are important in the analysis but can be predicted using ChPT from the pion
electromagnetic form-factor.~\cite{BT2} 
All parameters needed to determine $V_{us}$
can be determined from the scalar form factor in $K_{\ell3}$~\cite{BT2} and the
curvature can be predicted as well from knowledge about scalar form factors of
the pion.\cite{BD}

\section{Chiral Perturbation Theory}

Chiral Perturbation Theory is an effective low energy field theory which is
an approximation to Quantum Chromodynamics (QCD). 
It was introduced in its modern form by Weinberg,
Gasser and Leutwyler.~\cite{ChPT,GL1} 
 QCD in the limit of
massless quarks has a global $SU(3)_L\times SU(3)_R$ chiral symmetry. This
symmetry is spontaneously broken down to the vector subgroup
$SU(3)_V$ by the quark condensate being different from zero.
\be
\langle \bar q q\rangle = \langle \bar q_L q_R+\bar q_R q_L\rangle
\ne 0\,.
\ee
The eight broken generators lead to eight Goldstone bosons. These are massless
{\em and} their interactions vanish at zero momentum. This allows to build
up a well defined perturbative expansion in terms of momenta,
generically referred to as an expansion in $p^2$.
Quark masses can be counted as order $p^2$ since
$p_\pi^2 = m_\pi^2 \sim m_q \langle\bar q q\rangle$. Insertions
of external photons and $W^\pm$-bosons are counted as order $p$ since these
are incorporated via covariant derivatives. Recent lectures, providing many
more details than given here are available.~\cite{chptlectures}
One of the underlying problems is that ChPT is an effective field theory.
As such the number of parameters increases order by order. In ChPT in the
purely mesonic strong and semi-leptonic sector there are two
parameters at lowest order ($p^2$), ten at NLO ($p^4$),~\cite{GL1} and 90
at NNLO ($p^6$).~\cite{BCE1} The renormalization procedure and the divergences
are known in general to NNLO~\cite{BCE2} and provide a strong check on
all calculations. One problem in comparing different calculations
is the use of different renormalization schemes.
The calculations that were used to determine all the needed parameters
are those of the masses and decay
constants,~\cite{ABT1} $K_{\ell4}$~\cite{ABT2} and the
electromagnetic form factors.~\cite{BT1}

\section{$K_{\ell3}$ decays: definitions, $V_{us}$ and linearity of
the form factors}

There are two $K_{\ell3}$ decays:
\ba
{K_{\ell3}^+:} \hskip1cm
 K^+(p) &\rightarrow& \pi^0 (p') \ell^+ (p_\ell) \nu _\ell (p_\nu)
\nonumber
\\
{K_{\ell3}^0:} \hskip1cm
K^{0}(p) &\rightarrow &\pi^{-} (p') \ell^+ (p_\ell) \nu_\ell (p_\nu)\,.
\ea
The amplitudes for $K_{\ell3}^{+,0}$ can be written as
\ba
 T& =& \frac{G_F} {\sqrt{2}} V_{us}^\star \ell^\mu { F_\mu}^{+,0} (p',p)\,,
\quad\quad
{ F_\mu}^+ (p',p)
= \frac{1}{\sqrt{2}} [(p'+p)_\mu f^{K^+\pi^0}_+ (t) + (p-p')_\mu
f_-^{K^+\pi^0} (t)]
\nonumber\\
\nonumber \\
\ell^\mu &=& \bar{u} (p_\nu)\gamma^\mu  (1- \gamma_5) v (p_\ell)\,,
\quad\quad
 {F_\mu }^0 (p',p)
=  (p'+p)_\mu f^{K^0\pi^-}_+ (t) + (p-p')_\mu f_-^{K^0\pi^-} (t).
\ea
Isospin leads to the relations
\be
f^{K^0\pi^{-}}_+(t)=f^{K^+\pi^0}_+(t)= f_+(t)\quad\mbox{and}\quad
f^{K^0\pi^{-}}_- (t)=f^{K^+\pi^0}_- (t)= f_-(t)
\,,
\ee
We also define the scalar form factor and the usual linear parametrizations
\be
 f_0 (t) = f_+ (t) + \frac{t}{m^2_K - m^2_\pi} f_-(t)\,
\quad\quad
f_{+,0}(t) = 
f_+(0)\left(1+\lambda_{+,0}\frac{t}{m_\pi^2}\right)\,.
\ee
In order to determine $|V_{us}|$ we need to know theoretically 
and experimentally $f_+(0)$. On the theory side there are three
main effects. There is a well-known short-distance correction from
$G_\mu$ to $G_F$ calculated by Marciano and Sirlin. The corrections of
order $(m_s-\hat m)^2$ allowed by the Ademollo-Gatto theorem 
are estimated in the present work at $p^6$.
 The sizable isospin breaking discussed
by Leutwyler and Roos~\cite{LR} is in the process of being evaluated at order
$p^6$ too. On the experimental side, the old radiative correction calculations
used in Ref.~\cite{LR} have been updated in Ref.~\cite{Radiative} where
a clean procedure with generalized form-factors has been proposed.
The experimental data have so far been analyzed using a linear form factor
$f_+(t)$. The recent precise CPLEAR data~\cite{CPLEAR} are presented in a form
allowing to test this definition. Using a quadratic fit to their data
and {\em neglecting} systematic errors we get a normalized $f_+(0)=1$
and $\lambda_+=0.0245\pm0.0006$. Allowing for curvature we obtain
a sizable curvature and $f_+(0) = 1.008\pm0.009$ and
$\lambda_+=0.0181\pm0.0068$. The fitted curvature is compatible with zero
at the one sigma level and the central value is exactly at the ChPT prediction
given below. In order to obtain $|V_{us}|$ with an error of 1\% it is therefore
important to include the effect of curvature in the analysis.
Note that the central value of $\lambda_+$ is outside the errors quoted for the
linear fit.~\cite{CPLEAR}

\section{\boldmath$f_+(t)$: theory}

The ChPT calculation for $f_+(t)$ is rather long and cumbersome.
I will use here our work,~\cite{BT2} but an independent calculation
exists~\cite{Post3} and agrees reasonably
well.\footnote{The numerical disagreement
with some of the results mentioned there is under investigation.}
We can write the amplitude as
\ba
f_+(t) &=& 1 + f_+^{(4)}(t) + f_+^{(6)}(t)
\quad\mbox{with}\quad
f_+^{(4)}(t) = \frac{t}{2F_\pi^2}{ L_9^r}+\mbox{loops}\,,
\nonumber\\
f_+^{(6)}(t) &=& -\frac{8}{F_\pi^4}{ \left(C_{12}^r+C_{34}^r\right)}
\left(m_K^2-m_\pi^2\right)^2
+ \frac{t}{F_\pi^4} { R_{+1}^{K\pi}}
+ \frac{t^2}{F_\pi^4}  {(  - 4 C_{88}^r + 4 C_{90}^r )}
+\mbox{loops}(L_i^r)\,.
\ea
From the data on the pion electromagnetic form factor we~\cite{BT1}
obtain
\be
 L_9^r = 0.00593\pm 0.00043
\quad\mbox{and}\quad
  - 4 C_{88}^r + 4 C_{90}^r = 0.00022\pm0.00002\,.
\ee
Using these as input we can now fit the CPLEAR data and obtain
\begin{figure}
\begin{center}
\includegraphics[width=8.5cm,angle=-90]{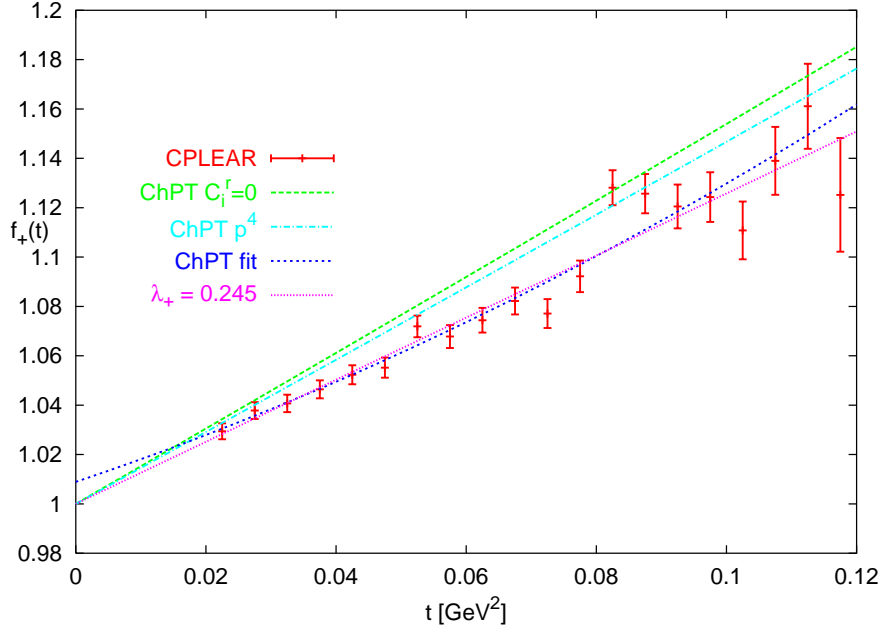}
\end{center}
\caption{ChPT fits to the CPLEAR data showing the effect of the
curvature compared to fits without it.
\label{fig:CPLEAR}}
\end{figure}
\be
{ R_{+1}^{K\pi}} =
 -(4.7\pm0.5)~10^{-5}~\mbox{GeV}^2\quad\mbox{and}\quad
 \lambda_+ = 0.0170\pm0.0015\,.
\ee
The first can be compared to the VMD estimate~\cite{BT2}
 ${R_{+1}^{K\pi}|_{VMD}} \approx -4~10^{-5}~\mbox{GeV}^2$.
The latter comes from ChPT as 
\be
\lambda_+ = 0.0283~(p^4)~~ + 0.0011~(\mbox{loops }p^6)~~ -0.0124 (C_i^r)\,. 
\ee

\section{\boldmath$f_0(t)$: theory}

The $p^6$ result for $f_0(t)$ can be rewritten as
\ba
f_0(t) &=& 1-\frac{8}{F_\pi^4}{\left(C_{12}^r+C_{34}^r\right)}
\left(m_K^2-m_\pi^2\right)^2
+8\frac{t}{F_\pi^4}{ \left(2C_{12}^r+C_{34}^r\right)}
\left(m_K^2+m_\pi^2\right)
\nonumber\\&&
+\frac{t}{m_K^2-m_\pi^2}\left(F_K/F_\pi-1\right)
-\frac{8}{F_\pi^4} t^2 { C_{12}^r}
+\overline\Delta(t)+\Delta(0)\,.
\ea
{\em 
$\overline\Delta(t)$ and
$\Delta(0)$ contain  NO $C_i^r$ and only depend on the
$L_i^r$ at order $p^6$ thus ALL
needed parameters can be determined experimentally.}
\be
\Delta(0) = -0.0080\pm0.0057[\mbox{loops}]\pm0.0028[L_i^r]\,.
\ee
is known and an expression for $\Delta(t)$ can be found in Ref.~\cite{BT2}
The errors are an estimate of higher orders and using fits of the $L_i^r$
using different assumptions.

\section{Conclusions}

I have discussed the calculation of the $K_{\ell3}$
form factors in ChPT at order $p^6$.
The main conclusions are that the curvatures for $f_+(t)$ and $f_0(t)$
can be predicted from ChPT and the data on pion electromagnetic~\cite{BT1}
 and scalar~\cite{BD}
form-factors, the curvature in $f_+(t)$ and $f_0(t)$
should be taken into account in new precision experiments but from the
slope and the curvature we can determine experimentally the needed
parameters to calculate $f_+(0)$. A precision of better than
one percent seems feasible for $|V_{us}|$.

\section*{Acknowledgments}
This work has been funded in part by
the Swedish Research Council and the European Union RTN
network, Contract No. HPRN-CT-2002-00311  (EURIDICE)

\end{document}